\documentclass[12pt,preprint]{emulateapj}

\newcommand{\um}{$\mu$m}

\newcommand{\kms}{km\thinspace s$^{-1}$}

\def\arcmin{\hbox{$^\prime$}}
\def\arcsec{\hbox{$^{\prime\prime}$}}
\def\utw{\smash{\rlap{\lower5pt\hbox{$\sim$}}}}
\def\udtw{\smash{\rlap{\lower6pt\hbox{$\approx$}}}}

\def\fha{\hbox{$^{\rm h}$}}
\def\fma{\hbox{$^{\rm m}$}}

\def\fdga{\hbox{$^\circ$}}
\def\farcma{\hbox{$^\prime$}}

\def\Lx{\hbox{\it L$_X$}}

\def\Msun{\hbox{\it M$_\odot$}}
\def\Minit{\hbox{\it M$_{\rm initial}$}}

\def\J{\hbox{\it J}}
\def\H{\hbox{\it H}}
\def\K{\hbox{\it K}}

\def\Mk{\hbox{\it M$_{\rm K}$}}
\newcommand{\Ks}{{\it K$_{\rm s}$}}
\newcommand{\Al}{{\it A$_\lambda$}}
\newcommand{\Aks}{{\it A$_{\rm K_{\rm s}}$}}

\newcommand{\Av}{{\it A$_{\rm V}$}}

\def\simgr{\mathrel{\hbox{\rlap{\hbox{\lower4pt\hbox{$\sim$}}}\hbox{$>$}}}}

\shorttitle{Discovery of a young massive stellar cluster}
\shortauthors{Messineo et al.}

\begin{document}


\title{Discovery of a young massive stellar cluster near 
 HESS J1813-178}


\author{Maria~Messineo\altaffilmark{1}, 
        Donald~F.~Figer\altaffilmark{1}, 
	Ben~Davies\altaffilmark{1},
	R.~Michael~Rich\altaffilmark{2},
	E.~Valenti\altaffilmark{3},
        R.P.~Kudritzki\altaffilmark{4}}

\email{messineo@cis.rit.edu}

\altaffiltext{1}{Chester F. Carlson Center for Imaging Science, Rochester Institute
   of Technology, 54 Lomb Memorial Drive, Rochester, NY 14623-5604, United
   States}
\altaffiltext{2}{Physics and Astronomy Building, 430 Portola Plaza, Box 951547, Department of Physics and Astronomy, 
University of California, Los Angeles, CA 90095-1547}
\altaffiltext{3}{European Southern Observatory, Alonso de Cordoba 3107, 
Santiago, Chile}
\altaffiltext{4}{Institute for Astronomy, University of Hawaii, 2680 
Woodlawn Drive, Honolulu, HI 96822}


%

\begin{abstract} 
We present the serendipitous discovery of a young stellar cluster in
the Galactic disk at $l$=12$^\circ$. Using Keck/NIRSPEC, we 
obtained high-- and low--resolution spectroscopy of several stars in the
cluster, and we  identified one red supergiant and two blue
supergiants. The radial velocity of the red supergiant provides a
kinematic cluster  distance of 4.7$\pm$0.4 kpc, implying luminosities
of the stars consistent with their spectral types. Together with the
known Wolf--Rayet star located 2\farcm4 from the cluster center, the
presence of the red supergiant and the blue supergiants suggests a
cluster age of $6-8$ Myr, and an initial mass of $\ga 2000$
\Msun. Several stars in the cluster are coincident with X--ray
sources, including the blue supergiants and the Wolf--Rayet star. This
is indicative of a high binary fraction, and is reminiscent of the
massive young cluster Westerlund 1. The cluster is coincident
with two supernova remnants, SNR G12.72$-$0.0 and G12.82$-$0.02, and
the highly magnetized pulsar associated with the TeV $\gamma$--ray source
HESS J1813$-$178.  The mixture of spectral types suggests
that the progenitors of these objects had initial masses of $20-30$
\Msun.
\end{abstract}





\section{Introduction} Young stellar clusters are important tools for
investigating the structure, the chemical enrichment, and the current
star formation of a galaxy.  Furthermore, they are natural
laboratories to study the evolution of massive stars.  Massive stars
explode as supernovae creating neutron stars and black-holes, and are
believed to be sources of $\gamma$--ray bursts, which are the most
energetic explosions in the Universe. Because of their short
lifespans, massive stars are rare and predominantly observed in young
(few Myrs) massive ($\sim 10^{4}$ \Msun) clusters.  Only a handful of
massive stellar clusters are known in the Galaxy: important examples
include Arches, Quintuplet, Westerlund 1, and Westerlund 2
\citep{figer08}.  As these clusters lie in the plane, their detection
is hampered both by interstellar extinction and by the relatively low
numbers of member stars. Indeed, the census is severely incompletely,
as demonstrated by the recent discovery of two other massive clusters
\citep[RSGC1 and RSGC2,][]{figer03,davies07}.

\begin{figure}[!] \begin{center}
\plotone{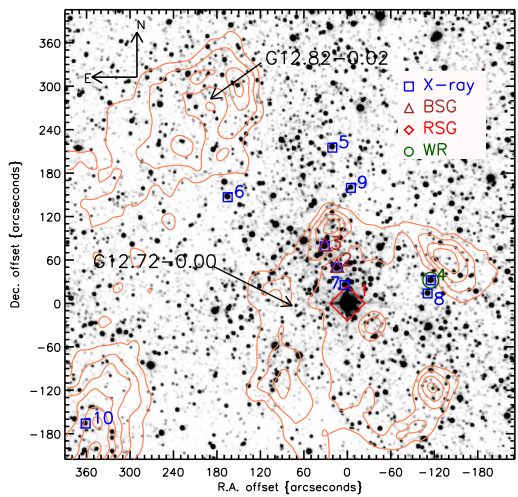}
\caption{\label{fig1}
\Ks--band image of the cluster from 2MASS, with overlayed contours of
90~cm data \citep{white05}. Contour levels, in
mJy beam$^{-1}$, are: 40, 60, 80, 100, 120, 140, 160, 200. The beam
size is 24\arcsec $\times$ 18\arcsec, FWHM, and the position angle of
the major axis is along the North--South direction.  Stellar
identification numbers are from Table~1.} 
\end{center}
\end{figure}

We have serendipitously discovered a new stellar cluster at RA= 18\fha
13\fma 24\fs31 and DEC=$-$17\fdga 53\farcma 30\farcs83
(J2000). Previously undetected, this cluster clearly appears as a
stellar over--density at infrared wavelengths (Fig.\ \ref{fig1}), and
is also visible at optical wavelength. It is located adjacent to the
star forming region W33 in an extremely rich and complex region. Two
supernovae remnants (SNRs), G12.82$-$0.02 and G12.72$-$0.0
\citep{brogan06}; a $\gamma$--ray source, HESS J1813$-$178
\citep{hess06}; and numerous X--ray sources \citep{funk07,helfand07}
have been detected in its surrounding area.

In this report, we present near--infrared spectra of four cluster
members, of which three are identified as early B--type stars, and one
as a red supergiant (RSG).  Also, we analyze existing photometric
infrared data and X--ray data, and find that several bright cluster
members are associated with an X--ray source, one of which is a known
Wolf--Rayet (WR) star \citep{hadfield07}.  From these results, we
estimate the age and distance of the cluster, which appears associated
with the star forming region W33 and the high--energy source HESS
J1813$-$178.

\section{Observations and data}   
Spectroscopic observations of several candidate cluster members (see
Table~1) were performed at the Keck Observatory on
2008 April 19 using NIRSPEC \citep{mclean98} under program U050NS
(P.I. R.M.  Rich). A spectrum of star \#1 (\Ks$=3.7$ mag) was obtained
with the NIRSPEC-7 filter and the 0\farcs57 $\times$ 24\arcsec\ slit,
covering $1.99-2.39$ $\mu$m at a resolution of R=17000 with two nodded
exposures of 10s each. Low--resolution spectra of sources \#2\ and
\#3\ were taken with the \K\ filter and a 0\farcs57 slit width,
covering $1.9-2.35$ $\mu$m at a resolution of R=1700.  Data reduction
was performed as described in \citet{figer03}. We subtracted pairs of
nodded frames and flat--fielded them.  The distorted 2-D spectral
traces were rectified onto a linear grid, using arc and etalon frames
for wavelength calibration. Atmospheric absorption and instrumental
response were removed by dividing each extracted target spectrum by
the spectrum of a B1V telluric standard (HD164581).

Photometric measurements of stellar point sources covering the cluster
region were obtained from the Vizier database.  We cross--correlated
2MASS near--infrared measurements \citep{cutri03}, Spitzer/GLIMPSE
mid--infrared data \citep{benjamin03}, as well as optical data from
the astrometric catalog NOMAD \citep{nomad}.

The region surrounding the cluster was observed with Chandra by
\citet{helfand07}, who detected  75 X--ray sources.  We
cross--identified the Chandra sources with the 2MASS and GLIMPSE
catalog using a radius of 1\farcs5, and found 44 matches.

\section{Spectral and Photometric Analysis}
\begin{figure}[!] 
\begin{centering}
\plotone{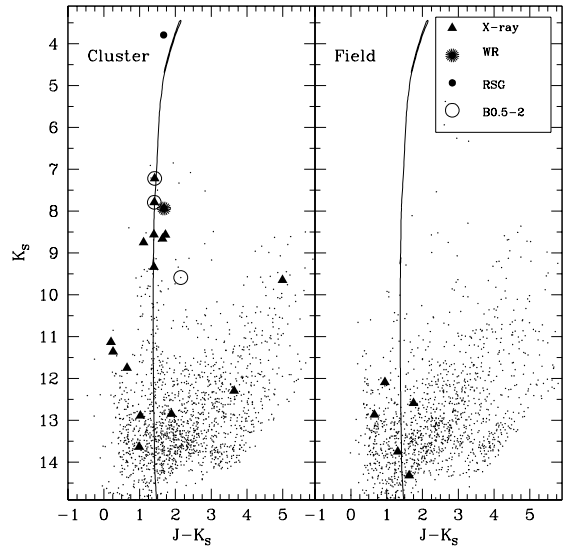}
\caption{\label{fig:cmd.ps} 2MASS \J$-$\Ks\ vs. \Ks\ CMDs.  In the
left panel, we show the CMD of point sources that are within
3\farcm5 from the center. In the right panel, field stars are
plotted, which were taken from an annulus of equal area at a radius of
5\arcmin. The vertical line indicates a solar isochrone of 6.3 Myr
\citep{lejeune01}, which was shifted to a reddening of
\Aks=0.83 mag and a distance of 4.7 kpc.  The starred symbol indicates
the location of the WR star, triangles the Chandra point
sources, the large dot the RSG star, and the circles the BSGs.}
\end{centering}
\end{figure}

In order to confirm that the observed over--density is an actual
stellar cluster, we analyzed the photometric properties of stellar
point sources from 2MASS and GLIMPSE surveys.  As an example, a (\J
$-$\Ks) vs \Ks\ diagram of 2MASS point sources within 3\farcm5 from
the peak of over--density is shown in Fig.\ \ref{fig:cmd.ps}. A well
defined cluster sequence appears in the infrared color--magnitude
diagram (CMD) at about \J $-$\Ks$=1.5$ mag and \Ks$=4-12$ mag. Its red
color and bright magnitude distinguish the cluster sequence from a
foreground component.
An interstellar extinction of \Aks$=0.83\pm0.2$ mag (Av$=9.1$
mag) is estimated by matching the colors of the observed cluster
sequence with a theoretical isochrone (6.3 Myr, solar composition)
from the Geneva group \citep{lejeune01}, and by assuming a power--law
extinction curve {\it \Al} $\propto \lambda^{-1.9}$ \citep{messineo05}.  
Because the main sequence is almost a vertical sequence when viewed 
in the near--infrared, this estimate is independent of age.

\begin{figure}[!]
\begin{center}
\plotone{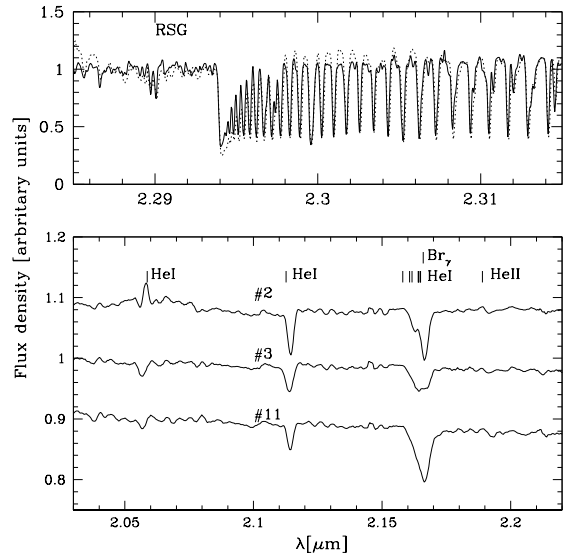}
\end{center}
\caption{\label{fig:spectra.ps} Top panel: high--resolution spectrum
of the RSG star in the region of the CO band--head feature at 2.293
$\mu$m. As a comparison, a spectrum of a K3 I star \citep{davies08} is
also shown (dashed line).  Bottom panel: low--resolution spectra of
\#2, \#3, and \#11. Helium and hydrogen lines
are taken from \citet{najarro94}.}
\end{figure}

\begin{deluxetable*}{rrrrrrrrrrrrrrrl}
\tablewidth{0pt}
\tablecaption{ List of candidate massive stars.}
\tablehead{
\colhead{ID}& 
\colhead{Ra}& 
\colhead{Dec}& 
\colhead{B} & 
\colhead{V} & 
\colhead{R} &
\colhead{J} & 
\colhead{H} & 
\colhead{K$_{\rm s}$} &
\colhead{[3.6]} & 
\colhead{[4.5]} &
\colhead{[5.8]} & 
\colhead{[8.0]} & 
\colhead{ID$_X$}      &			  
\colhead{\Lx}  & 
\colhead{Sp} 
}
\startdata
01 & 18:13:22.255 & $-$17:54:15.55 &  14.74 & 13.68 & 12.73 &	5.46 &  4.23 & 3.79 &	5.07 &  4.05&  3.57&\nodata&\nodata &\nodata& K2I\\      
02 & 18:13:23.256 & $-$17:53:26.54 &  20.08 & 15.03 & 14.50 &	8.64 &  7.72 & 7.22 &	6.94 &  6.78&  6.63&  6.68& 39 &  4.1&B0.5-2I\\
03 & 18:13:24.431 & $-$17:52:56.75 &  20.42 & 15.74 & 14.68 &	9.20 &  8.25 & 7.79 &	7.47 &  7.38&  7.30&  7.30& 43 &  4.8&B0.5-2I\\
04 & 18:13:14.188 & $-$17:53:43.60 &  20.48 & 17.57 & 15.20 &	9.62 &  8.60 & 7.94 &	7.34 &  6.93&  6.70&  6.35& 24 & 80.5&WN7b   \\
05 & 18:13:23.711 & $-$17:50:40.41 &  19.51 & 16.41 & 14.94 &	9.96 &  9.06 & 8.56 &	8.23 &  8.10&  7.99&  8.06& 41 & 63.4&\nodata \\
06 & 18:13:33.817 & $-$17:51:48.89 &\nodata &\nodata& 16.12 &  10.29 &  9.15 & 8.57 &	8.00 &  7.90&  7.75&  7.75& 58 &  4.0&\nodata\\
07 & 18:13:22.519 & $-$17:53:50.14 &  18.63 & 15.60 & 16.57 &  10.30 &  9.27 & 8.66 &	8.03 &  7.70&  7.49&  7.30& 37 & 10.3&\nodata\\
08 & 18:13:14.494 & $-$17:54:01.77 &  19.91 & 15.47 & 15.71 &	9.86 &  9.12 & 8.75 &	8.43 &  8.33&  8.31&  8.31& 27 &  3.8&\nodata\\
09 & 18:13:21.911 & $-$17:51:35.91 &  21.17 &\nodata& 16.24 &  10.74 &  9.80 & 9.34 &	8.95 &  8.90&  8.83&  8.91& 36 &  2.8&\nodata\\
10 & 18:13:47.561 & $-$17:57:01.43 &  15.32 &\nodata& 13.64 &  10.76 &  9.89 & 9.60 &	9.31 &  9.34&  9.22&  9.20& 71 & 21.0&\nodata\\
11 & 18:13:23.620 & $-$17:53:24.50 &\nodata &\nodata&\nodata&  11.74 & 10.85 & 9.59 &  10.16 & 10.32& 10.02&\nodata&\nodata&\nodata&B0.5-2\\
\enddata		
\tablecomments{For each star, number designations and coordinates
(J2000) are followed by magnitudes measured in different bands. \J,\H,
and \Ks\ measurements are from 2MASS, while the magnitudes at 3.6 \um,
4.5 \um, 5.8 \um, and 8 \um\ are from GLIMPSE.  $B$, $V$, and $R$
associations are taken from the astrometric catalog NOMAD. For sources
with  X--ray associations, number designations (ID$_X$) are taken
from \citet{helfand07}.  Estimates of \Lx\ are obtained for a distance
of 4.7 kpc, by assuming $N(H)= 1.6 \times 10^{22}$ cm$^{-2}$, a power 
law model, and a
photon index of 1.5 \citep{townsley06}.  X--ray luminosities \Lx\ are
given in $10^{31} erg~ s^{-1}$.}
\end{deluxetable*}

A variety of massive stellar objects was detected toward the new
stellar cluster: a RSG star, a WR star, several BSGs (Table~1). Star
\#4 is associated with a known WR star \citep[\#8 in~][]{hadfield07}.
Star \#1, a star with \Ks$=3.7$ mag, 3.5 mag brighter than other
cluster members, dominates the infrared cluster surface brightness.
Its high--resolution spectrum presents CO--bands in absorption as
shown in Fig.~\ref{fig:spectra.ps}.  A radial velocity
V$_{LSR}=+62\pm4$ \kms\ was measured by cross--correlating its
spectrum with that of Arcturus after re--binning the latter at the
same resolution.  A determination of the spectral type was obtained by
comparing the CO equivalent width with that of other template stars,
as explained in \citet{davies07}.  The resulting CO equivalent width
is consistent with that of a K2 I or an M4 III star.  From the
low--resolution frames, three spectra were extracted: the two spectra
of sources \#2\ and \#3, and a third spectrum of source \#11, which is
6\arcsec\ away from \#2, and fell on the slit.  The detection of HeI
lines, at 2.058 $\mu$m and 2.112 $\mu$m, and HI line at 2.166 $\mu$m,
and the lack of HeII indicate that \#2, \#3, and \#11 are early
B--type (B0-B2) stars \citep{hanson96}.  The emission line at 2.058
$\mu$m suggests that \#2 is a blue supergiant (BSG).

\section{Massive Stars and Distance} 
From the radial velocity of star \#1, we derive a kinematic
heliocentric distance of $4.7 \pm 0.4$ kpc by using the rotation curve
of \citet{brand93} and a solar Galactocentric distance of 7.6 kpc
\citep{kothes07}. Since the far distance  (10.3 kpc) would make the stars
over--luminous, we only consider the near distance. A
spectro--photometric distance of 0.9 kpc or 4.4 kpc is inferred when
assuming an M4 giant or an early K RSG, respectively, together with
the absolute infrared magnitudes of \citet{wainscoat92}.  An
interstellar extinction \Aks\ of $0.48$ mag is derived.  We conclude
that this bright star is a K2I RSG star. 
Star \#1 is 0\farcm9 from the center. 

By assuming a distance of 4.7 kpc, as calculated for the RSG star, an
intrinsic \H$-$\Ks\ from 0.11 to $0.27$ mag \citep{crowther06}, and
using the extinction law by \citet{messineo05}, we derive for the WN7b
star an \Aks\ from 0.83 to 0.59 mag and an absolute magnitude \Mk\ from
$-6.0$ to $-6.2$. This value of \Mk\ is consistent with those of
similar WN7 stars in Westerlund 1 \citep{crowther06}. The 2MASS colors
and magnitudes are consistent with being a cluster member. 
A bolometric correction of $-3.5$ \citep{crowther06} yields
a bolometric magnitude of $-9.6$.  
Star \#4 is 2\farcm4 from the center. 

The 2MASS photometric measurements of \#2 and \#3 support either early
nearby B dwarfs  or early B supergiants at a distance of  $3.7\pm1.7$
kpc.  We assume the absolute magnitudes and bolometric  corrections given
by \citet{crowther06b} and \citet{bibby08},  and consider both extinction
laws by \citet{messineo05} and  \cite{indebetouw05}. Star \#2\ has
\Aks$=0.87$ mag, and \#3 has \Aks$=0.81$ mag \citep{messineo05}. 
However,  they have the same interstellar extinction as the WR star; they
are  located in the same region of the CMDs, and most likely they are
B0--2  supergiants, as well as members of the cluster. They are within
0\farcm6 from the center. Star \#11 has a
poor  2MASS photometry; with \Ks$=9.59$ mag and an extinction \Aks$=1.2$
mag  it is consistent with being an early B dwarf.

We conclude from the CMDs and distance estimates, that the RSG, the WR
star, and the BSGs are all part of the same stellar cluster.
The average spectrophotometric distance of $3.7\pm1.7$ kpc is consistent
with the kinematic distance $4.7\pm0.4$ kpc within uncertainties.
We assume the kinematic distance.

\section{The X--Ray associations and energetics} 
The 2MASS CMD (Fig.\ \ref{fig:cmd.ps}) shows two distinct populations
of Chandra sources. One is associated with a blue foreground main
sequence population, while the other is associated with redder and
brighter stars that have colors and magnitudes consistent with a
cluster membership.  X--ray sources with a bright 2MASS counterpart
(\Ks$<9.6$ mag), such as candidate massive members of the stellar
cluster, are more spatially concentrated; except for one, they are
within 3\farcm5, while the other X--ray detections are within
12\arcmin\ from the cluster center.  The nine candidate massive
members with an X--ray association are included in Table~1.  Star \#4,
the WR star \citep[\#8 in~][]{hadfield07}, is associated with the
Chandra source \#24 \citep{helfand07}, coinciding with the XMM source
\#2 of \citet{funk07}.

Massive stars can emit X--rays. Single OB stars with shocked stellar
winds can  emit with a typical X--ray luminosity \Lx\ of 10$^{31-33}$ erg
s$^{-1}$ \citep{pollock87}. Shocks between the colliding winds of OB+OB
or OB+WR binaries can generate a \Lx\ of 10$^{32-35}$ erg s$^{-1}$
\citep{clark08}. 

To estimate the X--ray fluxes of the nine bright infrared stars
associated with an X--ray source, we used an interstellar extinction of
\Av$=9.1$ mag, which was estimated from the CMDs, as well as the
relationship between extinction and total hydrogen column density
$N(H)[\rm cm^{-2}] = A_V[\rm mag] \times 1.8 \times 10^{21}$.  Count
rates  were estimated from the Chandra/ACIS-I counts (in the band 2-10
kev)  that were given by \citet{helfand07}.  We converted them into
X--ray  unabsorbed fluxes by using the PIMMS v3.9d, an ACIS count
estimator,  with a power law model and a photon index of 1.5
\citep{townsley06}. Fluxes were converted into X--ray luminosities \Lx\
by assuming a distance of 4.7 kpc.  Six of the sources have \Lx$=2-10
\times 10^{31}$ erg s$^{-1}$.  The three remaining sources (\#4, \#5, and
\#10) are an order of magnitude brighter ($2-8 \times 10^{32}$ erg
s$^{-1}$); the brightest, \#4, is the WR star \citep{hadfield07}. The
ratio between the X--ray and bolometric luminosities was found to be $3.8
\times 10^{-7}$ for the WR star, but about 10 times fainter ($0.4 \times
10^{-7}$) for \#2\ and \#3, as expected for stars with spectral type
later than B1 \citep{cohen96,waldron07}. We conclude from the X--ray
energetics that these bright infrared stars are likely to be massive
stars. The WR is most likely a colliding wind binary \citep{portegies02}.

\section{Age and mass}  
The large variety of evolved objects -- 1 WR, 1 RSG, 2 BSGs, and
several X--ray emitters -- allows us to constrain the age and mass of
the stellar cluster by assuming coevality. The non--rotating Geneva
models with solar abundance predict the
onset of RSG stars at an approximate age greater than $6-7$ Myr
\citep{lejeune01}.  In a population of 6 Myr the
most massive stars have initial masses \Minit\ of approximately 29 \Msun\
\citep{lejeune01}.  In contrast, WR stars are
present in populations younger than 7.9 Myr (\Minit$ > 20$
\Msun). In particular, single WR stars have initial
masses greater than $26-30$ \Msun, whereas binary WR stars have
initial masses greater than $20-25$ \Msun\ \citep{eldridge08}. The
X--ray emission associated with our WR star suggests a binary system.
We conclude that the cluster is $6-8$ Myr old since this age allows
for the coexistence of both WR and RSG stars. Models with non--zero rotation  
increase both limits  by $\sim$20\% \citep{mey05}.

Assuming that the other eight X--ray emitters associated with the
cluster, other than the WR star, are BSGs with masses larger than 20
\Msun, and by assuming a Salpeter IMF down to 1.0~\Msun, we derive a
total initial cluster mass of 2000 \Msun. If we add as potential BSGs
24 other stars within 3\farcm5 of the center and with colors and
magnitudes similar to the X--ray sources (\Ks$<9.6$ mag, \J$-$\Ks$<3$
mag), the number of BSGs increases to 33, and the estimated initial
mass of the cluster to 6500 \Msun.

\section{Supernovae remnants} 
The cluster core is about 4\farcm5 away from HESS J1813$-$178. Several
studies have recently been carried out in order to unveil the nature
of the HESS $\gamma$--ray source. As a result, numerous radio and
high--energy sources have been detected (see Fig.\
\ref{fig1}).  Two non--thermal radio shells, G12.82$-$0.02 and
G12.72$-$0.00, were identified on the 90 cm VLA survey
\citep{brogan06}, a few arcminutes one from another (see Fig.\
\ref{fig1}).  Very little is known about the SNR
G12.72$-$0.00. The other SNR was associated with the HESS J1813$-$178
source by \citet{helfand04}.  A pulsar wind nebula (PWN) was detected
by Chandra within this SN radio shell less than 1\arcmin\ from the
maximum probability centroid of the HESS source
\citep{helfand07,funk07}.  By analyzing the X--ray flux, the authors
concluded that the SN G12.82$-$0.02, the HESS J1813$-$178 source, and
the PWN lie at or just beyond 4 kpc, and might be associated with the
star forming region W33. Similar conclusions were also obtained with
the XMM data by \citet{funk07}.  No stellar counterpart to the PWN was
found in the 2MASS or Spitzer/GLIMPSE images by
\citet{helfand07}. \citet{dean08} assumed a distance of 4.5 kpc and an
age of 300 yr, and found the putative pulsar to have an extremely high
magnetic field (B $= 1.28 \times 10^{14}$ G).  \citet{funk07} detected six other XMM sources in
addition to the PWN, AXJ1813$-$178, and \citet{helfand07} detected a
total of 75 Chandra sources in the region surrounding the two SNRs
(Fig.\ \ref{fig1}).

The new stellar cluster is coincident (to within 1\farcm6)  with the radio
shell of SN G12.72$-$0.00,  which suggests its association with the supernovae
progenitor. The kinematic cluster  distance of 4.7 kpc is consistent with the
distance to the high--energy source HESS J1813$-$178, as inferred from the
hydrogen column density by \citet{funk07} and \citet{helfand07}. Massive members
of this cluster were most likely the progenitors of the two supernovae and of
the  pulsar associated with HESS J1813$-$178.  The progenitors of these objects
had likely an initial mass similar to that of the RSG and WR stars ($20-30$
\Msun).

\acknowledgments { The material in this work is supported by NASA
under award NNG 05-GC37G, through the Long--Term Space Astrophysics
program. This research was performed in the Rochester Imaging Detector
Laboratory with support from a NYSTAR Faculty Development Program
grant.  Part of the data presented here was obtained at the
W. M. Keck Observatory, which is operated as a scientific partnership
among the California Institute of Technology, the University of
California, and the National Aeronautics and Space Administration. The
Observatory was made possible by the generous financial support of the
W. M. Keck Foundation. This publication makes use of data products
from the Two Micron All Sky Survey, which is a joint project of the
University of Massachusetts and the Infrared Processing and Analysis
Center/California Institute of Technology, funded by the National
Aeronautics and Space Administration and the NSF. This research has
made use of Spitzer's GLIMPSE survey data, the simbad and Vizier
database. RMR acknowledges support from grant AST-0709479
from the National Science Foundation.
}


\end{document}